\documentclass[reprint,amsmath,amssymb,aps,showpacs,prb,longbibliography]{revtex4-1}
\usepackage[utf8]{inputenc}

\usepackage{graphicx}
\usepackage{dcolumn}% Align table columns on decimal point
\usepackage{bm}% bold math
\usepackage[colorlinks=true]{hyperref}% add hypertext capabilities
\usepackage{mathrsfs}
\usepackage{amsthm}
\usepackage{bm}

\newcommand{\tr}{\text{tr}}
\renewcommand{\bf}[1]{\textbf{#1}}
\def\be{\begin{equation}}
\def\ee{\end{equation}}
\def\bea{\begin{eqnarray}}
\def\eea{\end{eqnarray}}

\usepackage{xcolor}

\begin{document}

\title{Localization anisotropy and complex geometry in two-dimensional insulators}

\author{Bruno Mera}
\email{bruno.mera@tecnico.ulisboa.pt}
\affiliation{Instituto de Telecomunica\c{c}\~oes, 1049-001 Lisboa, Portugal}
\affiliation{Departmento de F\'{i}sica, Instituto Superior T\'ecnico, Universidade de Lisboa, Av. Rovisco Pais, 1049-001 Lisboa, Portugal}

\date{\today}% It is always \today, today,

\begin{abstract}
The localization tensor is a measure of distinguishability between insulators and metals. This tensor is related to the quantum metric tensor associated with the occupied bands in momentum space. In two dimensions and in the thermodynamic limit, it defines a flat Riemannian metric over the twist-angle space, topologically a torus, which endows this space with a complex structure, described by a complex parameter $\tau$. It is shown that the latter is a physical observable related to the anisotropy of the system. The quantity $\tau$ and the Riemannian volume of the twist-angle space provide an invariant way to parametrize the flat quantum metric obtained in the thermodynamic limit. Moreover, if by changing the couplings of the theory, the system undergoes quantum phase transitions in which the gap closes, the complex structure $\tau$ is still well defined, although the metric diverges (metallic state), and it is fixed by the form of the Hamiltonian near the gap closing points. The Riemannian volume is responsible for the divergence of the metric at the phase transition.
\end{abstract}

\maketitle

%\tableofcontents

\section{Introduction}
\label{sec: Introduction}

Condensed-matter theory has been going through a paradigm shift. The notion of topological phases has been largely responsible for it. Systems of gapped free fermions that were thought to be completely understood from the traditional point of view, solvable through a Bogoliubov-Valatin transformation, have shown to be far from trivial and have a rich topological character. The integer quantum Hall effect is a paradigmatic example of the type of phenomena occurring in topological phases, the explanation of which is ultimately related to the topological non triviality of a vector bundle describing the twist of the occupied Bloch bands. Haldane's seminal paper~\cite{hal:88} provided the first instance of a Chern insulator, in which a quantized and topologically protected (provided the gap is maintained) transverse conductivity appears even though there is no net magnetic field. Topology is also responsible for the stability of the gapless edge modes appearing at the boundary of topological insulators and superconductors~\cite{ryu:hat:02, jac:12, kit:01}. Topological phases cannot be described through the Landau-Ginzburg theory associated with some local order parameter. Instead, they are effectively described through topological invariants, such as integrals of Chern classes of a vector bundle over a torus, an example of which is the Thouless--Kohomoto--Nightingale--den Nijs invariant~\cite{tknn:82}, or the holonomy of a flat connection, such as the Zak phase describing the polarization of a one-dimensional insulator~\cite{zak:89, ber:84}. The classification of symmetry protected topological phases of gapped free fermions~\cite{kit:09,sch:ryu:fur:lud:08, sch:ryu:fur:lud:09,kru:deB:van:kan:sla:17} can be described in terms of homotopy groups or \emph{K}-theory~\cite{kar:08}. This shows how sophisticated ideas from mathematics find a natural home in the study of phases of matter.

Finer properties of insulating states refer to their geometry and not just the topology~\cite{res:11, res:18}. The momentum space quantum metric is related to the overlap between Bloch states between adjacent momenta~\cite{mat:ryu:10, res:11, res:18}. It turns out that the integral over the momentum of this metric (see the Supplemental Material of Ref.~\cite{oza:gol:19} for a derivation) is naturally associated with the spread functional and the localization tensor of the material~\cite{mar:van:97, res:98, res:sor:99, sou:wil:mar:00, mat:ryu:10}. The localization tensor encodes how the electronic degrees of freedom are spatially organized, and it allows us to distinguish between insulators and metals. In particular, the localization tensor is naturally related to the conductivity of the material through fluctuation-dissipation relations~\cite{res:11}. Numerous proposals for extracting the quantum metric have been presented (see Refs.~\cite{kol:gri:pol:13, neu:cha:mud:13, oza:18, oza:gol:18, ble:sol:mal:18, ble:mal:gao:sol:18, kle:ras:cue:bel:18}). The first successful measurement of the quantum metric and, actually, the full quantum geometric tensor, whose real part is the quantum metric and whose imaginary part is the Berry curvature, was performed using nitrogen vacancy centers in diamond, see Ref.~\cite{yu:yan:gon:gol:19}. Later, using tunable superconducting circuits, the quantum metric was also measured~\cite{tan:zhang:yang:19}. A method on how to measure the localization tensor through spectroscopy in synthetic quantum matter (such as ultracold atomic gases or trapped ions), relying on the fluctuation-dissipation theorem, was recently proposed in Ref.~\cite{oza:gol:19}.

In the present paper, in the spirit of the geometrization of the theory of the insulating state of matter, we introduce a complex quantity $\tau$ which measures the anisotropy in localization of two-dimensional insulators. This quantity provides, as we will see, together with the Riemannian volume of the twist-angle space, an invariant way to parametrize the flat quantum metric obtained in the thermodynamic limit. Moreover, $\tau$ defines a complex structure in the torus of twist angles; that is, it gives it the structure of a complex one-dimensional manifold. Two different $\tau$'s related by modular transformations define isomorphic complex manifolds, and this will be shown to be related to choosing different lattice bases in real space; hence, they differ by a gauge choice and should be identified. If by changing the couplings, the system undergoes a phase transition, going through a metallic state by closing the gap, then the quantity $\tau$, unlike the volume, is still well defined in the thermodynamic limit. Moreover, we will see that its behaviour is tied to the low-energy theory near the gap closing points. The invariant description obtained is relevant to be able to compare different measurements since it makes no reference to gauge choices.

The paper is organized as follows. In Sec.~\ref{sec: The geometry of threading a flux through the system in the case of band insulators}, we discuss the geometry of threading a flux through the system in the case of band insulators. Afterwards, in Sec.~\ref{sec: The quantum metric in twist-angle space, the localization tensor and the complex structure tau}, we consider the quantum metric in twist-angle space, relate it to the localization tensor, introduce the complex structure $\tau$ and discuss gauge invariance. In Sec.~\ref{sec: Relation to the low-energy theory near a quantum phase transition and the geometric character of tau}, we discuss the relation of $\tau$ to the low energy theory near a quantum phase transition, discuss the geometric character of $\tau$, comparing its behaviour to that of the Berry curvature in twist-angle space which is topological in character. Subsequently, in Sec.~\ref{sec: Examples}, we provide the example of a modified massive Dirac model, which allows to explore all possible values of $\tau$, and the paradigmatic Haldane model. Finally, in Sec.~\ref{sec: Conclusions and outlook}, we present the conclusions and outlook.
\section{The geometry of threading a flux through the system in the case of band insulators}
\label{sec: The geometry of threading a flux through the system in the case of band insulators}

Within a tight-binding model of an insulator in two spatial dimensions, in the presence of translation invariance, the Hamiltonian is described by an $n\times n$ matrix $H(\bf{k})$ depending smoothly on the quasimomentum $\bf{k}$ in the Brillouin zone ($\mbox{BZ}$), topologically a two-torus. Here $n$ is the number of internal degrees of freedom, such as orbitals or pseudospin. In the presence of a gap below the Fermi level $E_F$, the projector onto the valence bands, assuming we have $r<n$ such bands, $P(\bf{k})=\Theta(E_F-H(\bf{k}))$ ($\Theta$ is the Heaviside step function) is smooth and, up to adiabatic deformation and in the absence of generic symmetries other than charge symmetry, determines the topology of the insulator. One can take a finite system with $N\times N$ sites and periodic boundary conditions. When one goes to momentum space, this amounts to sampling the matrix $H(\bf{k})$ at points of the form $\bf{k}=(2\pi/N)\bf{m}$, with $\bf{m}\in \{0,...,N-1\}^2$. The ground state of the insulator is obtained by occupying all the valence bands.

Threading a flux through the finite system with periodic boundary conditions means that the fermions will acquire phases $\exp(-i\phi_1)$ and $\exp(-i\phi_2)$ when they are moved adiabatically around the fundamental loops in position space. This corresponds to placing the system on a torus with twisted boundary conditions. For this reason the angles $\phi_1$ and $\phi_2$ are referred to as twist-angles~\cite{niu:tho:wu:85, kud:wat:kar:hat:19,oza:gol:19}. In momentum space, threading a flux through the finite system with periodic boundary conditions corresponds to sampling the matrix $H(\bf{k})$ at points of the form $\bf{k}=(2\pi/N)\bf{m}+\bm{\phi}/N$, with $\bf{m}\in\{0,...,N-1\}^2$. In Fock space, this produces a smooth family of ground states parametrized by the twist angles: a line bundle $\mathcal{L}$ over the twist-angle torus $T^2$. The Berry curvature of this family has the form (see Ref.~\cite{kud:wat:kar:hat:19})
\begin{align*}
\mathcal{F}(\bm{\phi})=\frac{1}{N^2}\sum_{\bf{m}}\tr\big[F_{12}(\frac{2\pi\bf{m}}{N}+\frac{\bm{\phi}}{N})\big]d\phi_1\wedge d\phi_2,
\end{align*}
where $F_{12}(\bf{k})=P(\bf{k})[\frac{\partial P}{\partial k_1}(\bf{k}),\frac{\partial P}{\partial k_2}(\bf{k})]P(\bf{k})$ is the Berry curvature of the occupied Bloch vector bundle $E\to \mbox{BZ}$, whose fiber at $\bf{k}\in \mbox{BZ}$ is the image of the projector $P(\bf{k})$.
In the thermodynamic limit $N\to\infty$, noticing that the previous expression is a Riemann sum, the Berry curvature is flat, i.e., independent of $\bm{\phi}$:
\begin{align}
\int_{\text{BZ}}\frac{d^2\bf{k}}{(2\pi)^2} \tr\big[F_{12}(\bf{k})\big]d\phi_1\wedge d\phi_2= \frac{\mbox{ch}_1}{2\pi i} d\phi_1\wedge d\phi_2,
\end{align}
where $\mbox{ch}_1$ is the first Chern number of the Bloch bundle of occupied states. In fact (see~\cite{niu:tho:wu:85, res:05, res:07, kud:wat:kar:hat:19}), $i e^2 \mathcal{F}_{12}(\bm{\phi})\equiv \sigma_{\text{Hall}}$ is the Hall conductivity of the insulator, where $e$ is the charge of the fermions. It follows that $\int_{T^2} i\mathcal{F}/2\pi=\mbox{ch}_1=\int_{\text{BZ}\cong T^2}\tr(iF)/2\pi$, which is the well-known result of Ref.~\cite{niu:tho:wu:85}. See also Witten's description of the microscopic and macroscopic Berry connections, corresponding to the momentum space and twist-angle space Berry connections, in Secs.~2.4 and 2.5 of Ref.~\cite{wit:16:lectures} and the note in~\footnote{Mathematically, this proves that, as a complex line bundle over the torus, $\mathcal{L}\to T^2$ is isomorphic to the line bundle $\det E\to \text{BZ}\cong T^2$, whose fiber over $\bf{k}\in \text{BZ}$ is the one-dimensional vector space of Slater determinants of $r$ valence band Bloch wave functions with momentum $\bf{k}$, with $r$ being the total number of valence bands.}.
\section{The quantum metric in twist-angle space, the localization tensor and the complex structure $\tau$}
\label{sec: The quantum metric in twist-angle space, the localization tensor and the complex structure tau}

We now consider the quantum metric in twist-angle space. It can be defined as the first nontrivial term in the expansion of the absolute value of the overlap between ground states with infinitesimally close twisted periodic boundary conditions. Formally, it is the pullback to the twist-angle torus of the Fubini-Study metric.  In a similar fashion to what happened in the case of Berry curvatures in twist-angle and momentum spaces, the quantum metric in twist-angle space is related to the quantum metric in momentum space by
\begin{align}
G(\bm{\phi})&=G^{\mu\nu}(\bm{\phi})d\phi_\mu d\phi_\nu\nonumber\\
&=\frac{1}{N^2}\sum_{\bf{m}}g^{\mu\nu}(\frac{2\pi\bf{m}}{N}+\frac{\bm{\phi}}{N})d\phi_\mu d\phi_{\nu},
\end{align}
with $g(\bf{k})=\tr(P dPdP)=g^{\mu\nu}(\bf{k})dk_{\mu}dk_{\nu}$ being the momentum space metric. In the thermodynamic limit $N\to \infty$, we obtain a flat metric on the twist-angle torus 
\begin{align}
G(\bm{\phi})=\int_{\text{BZ}}\frac{d^2\bf{k}}{(2\pi)^2}g^{\mu\nu}(\bf{k}) d\phi_\mu d\phi_{\nu}.
\end{align}
Moreover, this metric is nothing but the localization tensor describing the spread in real space of the ground state wave function~\cite{mat:ryu:10,res:11, oza:gol:19}:
\begin{align}
\langle X^{\mu}X^{\nu}\rangle -\langle X^{\mu}\rangle \langle X^{\nu}\rangle = \int_{\text{BZ}}\frac{d^2\bf{k}}{(2\pi)^2}g^{\mu\nu}(\bf{k})\equiv G^{\mu\nu},
\end{align}
where $X^\mu$ is the position operator. This also justifies our choice for contravariant indices on the metric tensor according to Einstein's conventions in position space. In fact, this makes explicit the covariance of $G$ with respect to lattice gauge transformations. Namely, when we prescribe position operators $X^{\mu}$ on the lattice, we have explicitly made an identification of the lattice with $\mathbb{Z}^2$ with a specific choice of a basis $e_{\mu}$, such that a lattice point is written as $X=X^{\mu}e_{\mu}$ for $X^{\mu}\in \mathbb{Z}$, $\mu=1,2$. This choice provides periodic (angular) coordinates on the Brillouin zone $k_{\mu}$ and on the twist-angle torus $\phi_\mu$ by writing $k=k_{\mu}e^{\mu}$ and $\phi=\phi_{\mu}e^{\mu}$ both modulo reciprocal lattice vectors $2\pi \mathbb{Z}e^1\oplus 2\pi\mathbb{Z}e^2$, where $e^{\mu}$ form a dual basis for $e_{\mu}$, i.e., $e^{\mu}(e_{\nu})=\delta^{\mu}_{\nu}$. It is now clear that we should identify quantum metrics over twist-angle space which differ by the action of $\mbox{GL}(2;\mathbb{Z})$ (invertible matrices of integers), since this would correspond to two observers that have chosen different basis vectors for the lattice and the physics should be ``gauge'' invariant. This action is simply, if we write $\widetilde{G}=[G^{\mu\nu}]_{1\leq \mu,\nu\leq 2}$,
\begin{align}
\widetilde{G}\mapsto A \widetilde{G} A^t
\end{align}
for $A\in \mbox{GL}(2;\mathbb{Z})$. Every metric on the torus $T^2$ is conformal to a flat metric like $G$. This a consequence of the uniformization theorem and the fact that conformal classes of metrics on $T^2$ are in one-to-one correspondence with complex structures (see, for example, Sec.~2.1. and Theorem~1.7 of Ref.~\cite{ima:tan:12}). The last statement is related to the fact that a conformal class specifies a way to measure angles between tangent vectors and hence, in two dimensions, uniquely specifies a 90º rotation on the tangent spaces, i.e., a complex structure. So, in principle, by changing the insulator, we would be able to move around the space of conformal classes of metrics or, equivalently, the space of complex structures. By a conformal class, we mean an equivalence class of metrics under the equivalence relation of conformal mappings (see Ref.~\cite{ima:tan:12}). To recover the conformal class of a flat metric we need also to identify flat metrics that differ by a scale transformation. Defining
\begin{align}
\tau = \frac{G^{12}}{G^{11}}+i\frac{\sqrt{\det \widetilde{G}}}{G^{11}}\in \mathcal{H}=\{\tau\in\mathbb{C}:\mbox{Im}(\tau)>0\},
\end{align}
the complex structure on $T^2$ is determined by the complex coordinate $z=\phi_1+\tau\phi_2$. The complex quantity $\tau$ together with the volume $V=\int_{T^2} \sqrt{\det (G)} d\phi_1 d\phi_2$ provides an invariant parametrization of all flat metrics in the twist-angle torus $T^2$, namely,
\begin{align}
G(\bm{\phi})=\frac{V}{(2\pi)^2\text{Im}(\tau)}[d\phi_1^2 +2\text{Re}(\tau)d\phi_1 d\phi_2 +|\tau|^2 d\phi_2^2].
\end{align}
The twist-angle space endowed with the flat quantum metric $G$ becomes a complex one-dimensional manifold, i.e., a Riemann surface~\cite{che:67, mir:95}. Moreover, we have to identify those $\tau$'s that differ by the induced action of $\mbox{SL}(2;\mathbb{Z})$ (since we fixed an orientation) by modular transformations,
\begin{align}
\tau\mapsto \frac{c+d\tau}{a+b\tau}, \ \left[\begin{array}{cc}
a & b \\
c & d
\end{array}\right]\in\mbox{SL}(2;\mathbb{Z}),
\end{align}
as these correspond to isomorphic complex manifolds (Theorem~1.1 of Ref.~\cite{ima:tan:12}). The quotient space $\mathcal{H}/\mbox{SL}(2;\mathbb{Z})$ has a fundamental domain given by $D=\{\tau\in\mathcal{H}: |\tau|\geq 1 \text{ and }  \mbox{Re}(z)\leq 1/2\}$. The points $\tau,\tau'\in D$ with $\mbox{Re}(\tau)=\pm 1/2$ and $\tau'=\tau\pm 1$ or $|\tau|=1$ and $\tau'=-1/\tau$ are the same in the quotient space.
Notice that the resulting complex twist-angle tori are to be understood as $\mathbb{C}/2\pi (\mathbb{Z}\oplus\tau\mathbb{Z})$. The choice of a basis on the lattice gave us a preferred dual basis and identified our torus with $\mathbb{R}^2/2\pi\mathbb{Z}^2$. The natural thing to do would then be to take $\tau=i$ and take a complex coordinate to be $z=\phi_1+i\phi_2$. However, the quantum metric can choose a different complex manifold structure on the twist-angle torus, and this structure is associated with the anisotropic spread of the ground state wave function in real space.

The localization tensor, or, equivalently, the quantum metric in twist-angle space $G$, is finite for an insulator, and it diverges when the system undergoes a quantum phase transition and becomes metallic~\cite{res:11}. The complex structure $\tau$ associated with $G$, however, is finite in both cases. Moreover, at the critical point, we will show it is determined uniquely by the low-energy properties of the theory and it is a measure of localization anisotropy of the system.
\section{Relation to the low-energy theory near a quantum phase transition and the geometric character of $\tau$}
\label{sec: Relation to the low-energy theory near a quantum phase transition and the geometric character of tau}

Suppose we are given a single-particle Hamiltonian and that two levels cross generically at a critical momentum $\bf{k}_c$ by tuning a parameter $M$ of the system to a value $M_c$. By a shift of the variables we can assume $\bf{k}_c=0$ and $M_c=0$. The two-level crossing can be described, in a neighborhood of $(\bf{k},M)=0$, by a $2\times 2$ low energy Hamiltonian of the form
\begin{align}
H(\bf{k},M)\approx (ak_1 +b k_2)\sigma_1 +(ck_1+dk_2)\sigma_2 +M\sigma_3,
\end{align}
with $a,b,c,d$ real parameters with $ad-bc\neq 0$, and $\sigma_{\mu}$, $\mu=1,2,3$, being the Pauli matrices. The new momentum variables $\bf{q}=(q_1,q_2)$ given by
\begin{align}
\left[\begin{array}{cc}
q_1\\
q_2
\end{array}\right]= \left[\begin{array}{cc}
a & b \\
c & d
\end{array}\right]\left[\begin{array}{cc}
k_1\\
k_2
\end{array}\right]
\end{align}
render this block an isotropic Dirac Hamiltonian,
\begin{align}
q_1\sigma_1+ q_2\sigma_2 +M\sigma_3.
\label{eq:isotropic Dirac Hamiltonian}
\end{align}
Notice that by making this change of variables with arbitrary real parameters we are explicitly violating the dual basis provided initially; that is, the new coordinates are not periodic coordinates on the torus except when 
\begin{align*}
A=\left[\begin{array}{cc}
a & b \\
c & d
\end{array}\right]\in \mbox{GL}(2;\mathbb{Z}).
\end{align*}
In these new coordinates, the momentum space quantum metric receives a contribution of the form
\begin{align}
\left[\begin{array}{cc}
\frac{q_1^2 + M^2}{(q_1^2+q_2^2+M^2)^2} & -\frac{q_1 q_2}{(q_1^2+q_2^2+M^2)^2}\\
-\frac{q_1 q_2}{(q_1^2+q_2^2+M^2)^2} & \frac{q_2^2 + M^2}{(q_1^2+q_2^2+M^2)^2}
\end{array}\right] + \text{regular},
\label{eq: momentum qmetric singular}
\end{align}
where regular means possible additional terms which are smooth in the limit $M\to 0$. Notice that as $\bf{q}\to 0$ and $M\to 0$ the contribution is singular since it goes as $(1/M^2) I_2$, where $I_2$ is the $2\times 2$ identity matrix. We can then assume that as $M\to 0$, the main contribution to the quantum metric is given by an arbitrarily small neighborhood of the critical point $\bf{q}=0$. The quantity of interest is therefore the integral
\begin{align}
\int_{\{\bf{q}:|\bf{q}|<\Lambda\}}\frac{d^2\bf{q}}{(2\pi)^2}\left[\begin{array}{cc}
\frac{q_1^2 + M^2}{(q_1^2+q_2^2+M^2)^2} & -\frac{q_1 q_2}{(q_1^2+q_2^2+M^2)^2}\\
-\frac{q_1 q_2}{(q_1^2+q_2^2+M^2)^2} & \frac{q_y^2 + M^2}{(q_x^2+q_y^2+M^2)^2}
\end{array}\right]
\end{align}
for some cutoff $\Lambda>0$. This integral can be evaluated explicitly, yielding
\begin{align}
\frac{1}{8\pi}\Big[\frac{\Lambda^2}{M^2+\Lambda^2}+\ln\big(\frac{M^2+\Lambda^2}{M^2}\big)\Big]I_2.
\end{align}
The first term is regular, but the second provides a logarithmic divergence. This is expected because the system becomes metallic as $M\to 0$. Had we performed the integration on the original $\bf{k}$ coordinates, since the transformation is linear, we would simply get, modulo multiplication by a positive constant, the matrix
\begin{align}
&\frac{1}{8\pi}\ln\big(\frac{M^2+\Lambda^2}{M^2}\big) A^tA +\text{regular} \nonumber \\
&=\frac{1}{8\pi}\ln\big(\frac{M^2+\Lambda^2}{M^2}\big)\left[\begin{array}{cc}
a^2+c^2 & ab+cd\\
ab+cd & b^2 +d^2
\end{array}\right]+\text{regular}.
\end{align}
This means that the quantum metric $G$, or equivalently, the associated matrix $\widetilde{G}$, reads
\begin{align}
\widetilde{G}= C\ln\big(\frac{M^2+\Lambda^2}{M^2}\big)\left[\begin{array}{cc}
a^2+c^2 & ab+cd\\
ab+cd & b^2 +d^2
\end{array}\right] +\text{regular}
\label{eq: twist-angle qmetric}
\end{align}
for some constant $C>0$. Since $\tau$ is a ratio of matrix elements of $\widetilde{G}$, the regular part does not contribute as it can be made zero in the limit of $M\to 0$ by a conformal transformation. Thus, the singular part determines $\tau$:
\begin{align}
\tau= \frac{ab +cd +i|\det A|}{a^2+c^2}.
\end{align}
Notice that $\tau$ is just a function of $A$, which determines the anisotropy of the low-energy Dirac Hamiltonian. The columns of the matrix $A$ determine basis elements for the lattice corresponding to the complex manifold determined by $\tau$. Actually, assuming $\det A>0$, if we write $\omega_1=a+ic$ and $\omega_2=b+id$, then $\tau=\omega_2/\omega_1$. We also remark that the volume goes as $V\sim \ln\big[(M^2+\Lambda^2)/M^2\big]$. Therefore, it is $V$ that is responsible for the singularity in the localization tensor in the thermodynamic limit.

Notice that if $A=I_2$, the isotropic Dirac Hamiltonian, then $\tau=i$, and $z=\phi_1+i\phi_2$, i.e., the usual complex structure as determined by the canonical basis. Moreover, if $A\in\mbox{SL}(2;\mathbb{Z})$, then $\tau\sim i$, so it is isomorphic to the isotropic case. This means that the $\bf{q}$ coordinates are again periodic coordinates on the torus.

The situation described above is generic since the variety of Hermitian matrices with  at least one repeated eigenvalue has codimension $3$ in the manifold of $n\times n$ Hermitian matrices~\cite{neu:wig:00, arn:95}. One could imagine, for example, having a quadratic band crossing, but this situation, in the absence of any particular symmetry enforcing it, is adiabatically connected to the previous case with two gap closing points, and the resulting $\tau$ can be calculated by summing the two contributions and taking the limit in which the two gap points collapse into a single one. 

The same type of argument is used to prove that the change in the Chern number at a band crossing depends uniquely on the details of the low energy Dirac theory near the critical point (see Refs.~\cite{sim:83,bel:95, arn:95}). If there is more than a single critical $\bf{k}_c$ for which the gap closes (and this has to happen for a finite number of them since the torus is a compact two-dimensional manifold), then we have to sum the individual contributions from each critical momentum to the quantum metric. Additionally, if we remain close to $M_c$, $\tau$ will be close to the one determined from the low-energy theory.

We remark that since $\tau$ is geometric and not topological, it is insensitive to gap inversions, i.e., to transitions in which the gap changes sign. This can be shown explicitly as follows. By looking at Eq.~\eqref{eq:isotropic Dirac Hamiltonian}, the situation of gap inversion corresponds to $M$ changing sign. From Eq.~\eqref{eq: twist-angle qmetric}, we see that since the singular part of $\widetilde{G}$ is insensitive to $M\to -M$, so is $\tau$. The same does not happen for the Berry curvature. Indeed, the contribution at momentum $\bf{q}=(q_1,q_2)$ to the integral yielding the Berry curvature in twist-angle space is
\begin{align*}
-\frac{i}{2}\frac{M}{\big(M^2+\bf{q}^2\big)^{3/2}},
\end{align*}
which is odd under $M\to-M$. This equation is to be compared with the analogous one for the quantum metric in twist-angle space, namely, Eq.~\eqref{eq: momentum qmetric singular}.

We would like to point out that the space of complex structures on the torus is an orbifold with orbifold points $\tau=i$ and $\tau=e^{i2\pi/3}$. These values of $\tau$ correspond to metrics with larger isometry groups. Generically, for a given $\tau$, the only nontrivial isometry allowed is $z\mapsto -z$, yielding a group $\mathbb{Z}_2$. However, $\tau=i$ is fixed when we take $\tau\mapsto -1/\tau$, yielding a $\mathbb{Z}_4$ isometry group. Similar reasoning for $\tau=e^{2\pi i/3}$ yields a $\mathbb{Z}_6$ isometry group. These two situations correspond to
\begin{align}
\tau=i  &\longleftrightarrow \widetilde{G}\sim\left[\begin{array}{cc}
1 & 0\\
0 & 1
\end{array}\right],\nonumber\\
\tau=e^{2\pi i/3} &\longleftrightarrow \widetilde{G}\sim \left[\begin{array}{cc}
1 & -1/2\\
-1/2 & 1
\end{array}\right]. 
\end{align}
The first case, as we will see below, is captured by the fully isotropic Dirac low-energy Hamiltonian, which in terms of the localization tensor implies that the $x$ and $y$ directions decouple since the off-diagonal terms are zero. The second case corresponds to a case where the $x$ and $y$ directions are cross correlated. This cross correlation is maximal in the sense that this corresponds to a boundary point of the fundamental domain $D$ with maximal $\mbox{Re}(\tau)\propto G^{12}$.
The isotropic case at the point of phase transition is very common in the models found in the literature, such as the massive Dirac model for a Chern insulator (see for example, Ref.~\cite{mat:ryu:10}) and the Haldane model~\cite{hal:88}, as shown below.
\section{Examples}
\label{sec: Examples}
\subsection{Modified massive Dirac model}
\label{subsec: Modified massive Dirac model}

A slight modification of the massive Dirac model given by
\begin{align}
H(\bf{k},M)&=[\sin(k_1)+a\sin(k_2)]\sigma_1+b\sin(k_2)\sigma_2 \nonumber \\
&+[M-\cos(k_1)-\cos(k_2)]\sigma_3,
\end{align}
with $a\in \mathbb{R}$ and $b>0$, will allow us to explore the space of complex structures of the twist-angle torus. Notice that the usual fully isotropic massive Dirac model is recovered for $a=0$ and $b=1$. The phase diagram stays identical, namely, there are phase transitions at $M=-2,0,+2$. The first Chern number of the Bloch bundle is $0$ for $|M|>2$, $+1$ for $-2<M<0$, and $-1$ for $0<M<2$. The gap closes at inversion-symmetric points, i.e., $\bf{k}=-\bf{k}$ modulo reciprocal lattice vectors. The low energy theories are given by the following:
\begin{itemize}
\item[(i)]Near $\bf{k}_c=(0,0)$,
\begin{align}
  H(\bf{k},M)=(k_1+ak_2)\sigma_1+b k_2\sigma_2 +(M-2)\sigma_3.
\end{align}
\item[(ii)]Near $\bf{k}_c=(0,\pi)$,
\begin{align}
  H(\bf{k},M)=(k_1-ak_2)\sigma_1-b k_2\sigma_2 +M\sigma_3.
\end{align}
\item[(iii)]Near $\bf{k}_c=(\pi,0)$,
\begin{align}
  H(\bf{k},M)=(-k_1+ak_2)\sigma_1+b k_2\sigma_2 +M\sigma_3.
\end{align}
\item[(iv)]Near $\bf{k}_c=(\pi,\pi)$,
\begin{align}
  H(\bf{k},M)=(-k_1-ak_2)\sigma_1-b k_2\sigma_2 +(M+2)\sigma_3.
\end{align}
\end{itemize}
In the above equations, $k_1$ and $k_2$ are understood as $k_1-k_{1,c}$ and $k_2-k_{2,c}$, respectively. From the prescription given before we have $\tau=a+ib$ for $M_c=\pm 2$ and $\tau=-a+ib$ for $M_c=0$ (both critical momenta contribute the same to the metric in this case). Notice that the usual Dirac model, i.e., for $a=0$ and $b=1$, has $\tau=i$ for all critical points. We see that the modified massive Dirac model allows us to explore all possible values of $\tau$.
\subsection{Haldane model}
\label{subsec: Haldane model}

For the Haldane model, one can show that the low-energy theory near the points $\bf{K}$ and $\bf{K}'$ has the following form:
\begin{itemize}
\item[(i)]Near $\bf{k}_c=\bf{K}$,
\begin{align}
  H(\bf{k})&=-\frac{3}{2}t_1(\frac{\sqrt{3}}{2}k_1+\frac{1}{2}k_2)\sigma_1+\frac{3}{2}t_1(\frac{1}{2}k_1-\frac{\sqrt{3}}{2}k_2)\sigma_2\nonumber\\
  &+M(\bf{K})\sigma_3.
\end{align}
\item[(ii)]Near $\bf{k}_c=\bf{K}'$,
\begin{align}
  H(\bf{k})&=\frac{3}{2}t_1(\frac{\sqrt{3}}{2}k_1+\frac{1}{2}k_2)\sigma_1+\frac{3}{2}t_1(\frac{1}{2}k_1-\frac{\sqrt{3}}{2}k_2)\sigma_2 \nonumber\\
  &+M(\bf{K}')\sigma_3.
\end{align}
\end{itemize}
In the previous formulas $t_1$ is a hopping amplitude, and $M(\bf{K})$ and $M(\bf{K}')$ are constant mass terms which depend on other couplings of the model.
Since $A=\left[\begin{array}{cc}
-\frac{\sqrt{3}}{2} & -\frac{1}{2}\\
\frac{1}{2} & -\frac{\sqrt{3}}{2} 
\end{array}\right]$ is a rotation matrix, it follows that $\tau=i$ at the critical points.
\section{Conclusions and outlook}
\label{sec: Conclusions and outlook}

Finally, we would like to point out that $\tau$ is a geometric quantity and not topological. Because of this, it is sensitive to adiabatic perturbations. However, at the critical points of phase transition it is determined, as shown above, by the low energy theory. Since symmetries constrain the low-energy theory, they also constrain $\tau$. In this sense, a future possible direction is to study the dependence of $\tau$ on generic symmetries. We remark that $\tau$ can be defined in the presence of interactions, provided some gap condition exists and we are given a nontrivial family of ground states of a problem parametrized by the two-torus of twist angles (more generally, any two-torus would do, but different physical interpretations for $\tau$ will appear). The family cannot be trivial since if it were trivial, i.e., constant, the metric would be automatically degenerate. Notice, however, that in the interacting case $G$ will not, generically, be flat. The procedure to determine $\tau$ is to determine a flat metric in the conformal class of $G$, which involves solving a differential equation for the conformal factor which enforces the Ricci scalar to be zero (see for example Sec.~1.3.2 of Ref.~\cite{ton:09}). Note also that $\tau$ and $V$ will no longer, in general, completely specify the localization tensor since it will not be flat. In the presence of translation invariance, this can also be seen as a measure of how interacting the system is. Namely, the failure of describing the localization tensor completely through $\tau$ and $V$ measures the fluctuations from a quasi-free-fermion description.

In summary, we have shown how in two dimensions the anisotropy of the localization tensor for band insulators is related to a complex structure over the twist-angle torus and that this quantity is finite, even when undergoing a phase transition in which a generic gap closing occurs, and thus going through a metallic state and intimately related to the anisotropy of the low-energy Dirac theory near the critical points.
The complex structure $\tau$ and the Riemannian volume $V$ are physically sensible gauge-invariant observables which completely characterize the localization tensor. The latter can be measured by means of spectroscopy in engineered quantum systems as proposed in Ref.~\cite{oza:gol:19}. 
\section*{Acknowledgements}
\label{sec: Acknowledgements}

B.M. is very grateful to N. Goldman and R. Resta for carefully reading, pointing out additional references for, and suggesting modifications to the manuscript. B.M. acknowledges very stimulating discussions with J. P. Nunes, J. M. Mourão, T. Baier, A. Carollo and N. Paunkovi\'{c}. B.M. is thankful for the support from SQIG -- Security and Quantum Information Group, under the Funda\c{c}\~ao para a Ci\^{e}ncia e a Tecnologia (FCT) project UID/EEA/50008/2020, and European funds, namely, H2020 project SPARTA. B.M. acknowledges projects QuantMining POCI-01-0145-FEDER-031826, PREDICT PTDC/CCI-CIF/29877/2017 and an internal IT project, QBigData PEst-OE/EEI/LA0008/2013, funded by FCT.

%%%%%
%\bibliographystyle{unsrt}
\bibliography{bib}
\end{document}